\documentclass[final,5p,times]{elsarticle}
\usepackage{amssymb,amssymb,graphicx}
\usepackage{graphicx}
\usepackage{booktabs}
\journal{Journal of Alloys and Compounds}
\newcommand{\p}{$\%$}
\newcommand{\pat}{{${~at.}$}\%}

\newcommand{\pn}{$\mathrm{R{_{N_2}}}$}

\newcommand{\tfn}{$\mathrm{Fe_{4}N}$}
\newcommand{\tcn}{$\mathrm{Co_{4}N}$}

\newcommand{\Ts}{$\mathrm{T_{s}}$}

\begin{document}

\title{Structural and magnetic properties of Co-N thin films deposited using magnetron sputtering at 523\,K}

\author {Nidhi Pandey$^1$, Mukul Gupta$^{1*}$, Rachana Gupta$^2$, Sujay Chakravarty$^3$, Neeraj Shukla$^3$, Anton Devishvili$^4$}
\address{$^1$UGC-DAE Consortium for Scientific Research, University Campus, Khandwa
Road, Indore 452 001, India}
\address{$^2$Institute of Engineering and Technology DAVV, Khandwa Road, Indore 452 017,
India}
\address{$^3$UGC-DAE Consortium for Scientific Research Kalpakkam Node,
Kokilamedu 603 104, Tamilnadu, India }
\address{$^4$Institut Laue-Langevin, rue des Martyrs, 38042 Grenoble
Cedex, France}

\address{$^*$ Corresponding author Email: mgupta@csr.res.in / dr.mukul.gupta@gmail.com/ \\Tel.: +91 731 246 3913, Fax: +91 731 246 5437}

\date{\today}


\begin{abstract}

In this work, we studied cobalt nitride (Co-N) thin films
deposited using a dc magnetron sputtering method at a substrate
temperature (\Ts) of 523\,K. We find that independent of the
reactive gas flow (\pn) used during sputtering, the phases of Co-N
formed at this temperature seems to be identical having N
\pat~$\sim$5. This is contrary to Co-N phases formed at lower \Ts.
For \Ts$\sim$300\,K, an evolution of Co-N phases starting from
Co(N)$\rightarrow$\tcn$\rightarrow$Co$_3$N$\rightarrow$CoN can be
seen as \pn increases to 100\p, whereas when the substrate
temperature increases to 523\,K, the phase formed is a mixture of
Co and \tcn, independent of the {\pn} used during sputtering. We
used x-ray diffraction (XRD) to probe long range ordering, x-ray
absorption spectroscopy (XAS) at Co absorption edge for the local
structure, Magneto-optical Kerr e ffect (MOKE) and polarized
neutron reflectivity (PNR) to measure the magnetization of
samples. Quantification of N \pat~was done using secondary ion
mass spectroscopy (SIMS). Measurements suggest that the magnetic
moment of Co-N samples deposited at 523\,K is slightly higher than
the bulk Co moment and does not get affected with the \pn~used for
reactive sputtering. Our results provide an important insight
about the phase formation of Co-N thin films which is discussed in
this work.

\end{abstract}


\maketitle

\section{Introduction} \label{intro}
Tetra 3d magnetic transition metal nitrides (e.g. \tfn, \tcn) are
ferromagnetic materials having higher (than pure metal) magnetic
moment, large spin polarization ratio (SPR) and superior chemical
stability. Such properties make them a
candidate in high density magnetic memory devices and
spintronics~\cite{PRB07_Matar, JAP14_Ito, 2011_Co4N_K_Ito,
CoN_AIP_Adv2015, JAC14_Lourenco, TSF14_Silva}. In addition, transition metal nitrides
are also used as an anode materials for
lithium-ion batteries. In particularly, cobalt nitrides find place in Li$_\mathrm{3-x}$Co$_{\mathrm{x}}$N, which shows a remarkable
high reversible capacity and good cycle performance~\cite{SSI00_Takeda_CoN_LIB, SSI09_DAS_CoN_LIB,JMC12_DAS_CoNLIB, RSC13_DAS_NiCoN_LIB}.
A lot of theoretical~\cite{JMMM99_P_Mohn_Matar, Coey.JMMM.1999,
JMMM92_Kuhnen, JMMM99_Sifkovits, JdP88_Matar} and
experimental~\cite{JMMM15_Dirba, MRB15_Li,
Atiq:JACS2009,APL08_Atiq, JCG15_Wang, APL11_Ito_Fe4N,
JAP15_KIto_Fe4N, Gallego.PRB04} reports are available on iron
nitride (Fe-N) system. In a number of studies \tfn~thin films have
been prepared and studied~\cite{JMMM15_Dirba, MRB15_Li,
Atiq:JACS2009,APL08_Atiq, JCG15_Wang,APL11_Ito_Fe4N,
JAP15_KIto_Fe4N, Gallego.PRB04}. On the other hand, the Co-N
system has not been explored as much. In particularly, recent
works on the \tcn~phase revealed that the SPR of \tcn~can be as
high as 0.9 (probably the highest) which is considerably larger
than the SPR of \tfn~at 0.66~\cite{Silva2015}. This has led to
renowned research on \tcn~phase during last couple of years.
Special interest has been paid to \tcn~thin films and they have
been deposited using sputtering~\cite{JMS87_Oda, Silva2015,
CoN_AIP_Adv2015, TSF14_Silva}, MBE~\cite{JCS11_Ito,
APL_11_K_Ito_Co4N} etc.

One of the key parameters for growth of single phase tetra metal
nitrides is the substrate temperature (\Ts). In the case of Fe-N,
the phase diagram is well-known and the \tfn~phase is formed at
\Ts $\sim$ 623\,K. In the absence of Co-N phase diagram, the
experimental method adopted for preparation of \tcn~thin film
seems to be influenced directly by the recipe used for \tfn~thin
films. However there seems to be large variations in \Ts~used for
preparation of \tcn~thin films; \Ts~as large as 723\, \Ts~as
low as 300\,K~\cite{JAP14_Ito, 2011_Co4N_K_Ito, JAC14_Lourenco, TSF14_Silva, APL_11_K_Ito_Co4N}.
In a recent work, It was demonstrated that
\tcn~thin film with lattice parameter close to its theoretical
values can be prepared at 300\,K~and these films were not stable
beyond 473\,K~\cite{APL_11_K_Ito_Co4N}. In an early study by Maya et. al.~\cite{JAP96_Maya_CoN}, thermal decomposition
of CoN thin films was measured. Here it was found that a volatile evolution of nitrogen
peaks around 613\,K. This signifies that N out diffuses from the CoN system leaving behind fcc Co. In the view of
such observations, higher \Ts ($\geq$600\,K) used for the growth of \tcn~phase seems
to be not appropriate. In order to clarify the phase formation of
\tcn~films we have chosen \Ts~at 523\,K, well-below the volatile evolution temperature.
In a recent study, Silva $et. al.$~\cite{Silva2015} also studied the formation of \tcn~thin films at \Ts=523\,K,
however they varied the partial nitrogen gas flow only in a narrow range. In the present work,
we did a systematic study by
preparing a series of Co-N thin films. We varied the partial nitrogen
gas flow in the whole range (0 to 100{\p}) and resultant films were
studied for their long and short range ordering, chemical composition
and magnetic properties. We found that the formation of Co-N phases at
\Ts = 523\,K~does not have any dependence on the partial nitrogen gas
flow. This is a new result for Co-N system and can be understood
in terms of its heat of formation.

\section{Experimental Procedure} \label{expe}

%\subsection{experimental Procedure, experimental measurements}

Co-N thin films were deposited on glass substrate at \Ts = 523\,K
by direct current magnetron sputtering (dcMS) using a AJA Int.
Inc. make ATC Orion-8 series sputtering system equipped with high
purity 3\,inch diameter cobalt target (99.99{\p}). The substrate
to target distance was fixed at 12\,cm. The substrate holder was
rotated along its own axis at 60 rpm during the deposition for
better uniformity of the film. The sputtering was done in Ar
(99.999\p) and N$_2$ (99.999\p) plasma environment using different
partial gas flows defined as {\pn} =
p$_{\mathrm{N}_2}$/(p$_{\mathrm{Ar}}$+p$_{\mathrm{N}_2}$, where
p$_{\mathrm{Ar}}$ and p$_{\mathrm{N}_2}$ are gas flow of Ar and
N$_2$ gases, respectively. With a base pressure of
1$\times$10$^{-7}$\,Torr, the pressure during deposition was kept
fixed at 3$\times$10$^{-3}$\,Torr using a dynamic throttle gate
valve. All samples were deposited at a fixed power of 100\,W and
\pn~used during deposition was 0, 25, 50, 75, 100{\p}. A reference
sample of pure Co was also deposited under identical conditions at
\Ts = 300\,K. The thickness of thin film samples was kept typically at 200\,nm, expect for \pn = 100\p,
sample, where the thickness was about 100\,nm due to reduced deposition rate.

To measure the concentration of nitrogen content, secondary ion
mass spectroscopy (SIMS) measurements were performed under UHV
conditions using a Hiden Analytical SIMS workstation. The crystal
structure and the phase formation of samples were characterized by
x-ray diffraction (XRD) using a standard XRD system (Bruker D8
Advance) using CuK$_{\alpha}$ x-ray source. To get the precise
information about the electronic structure, soft x-ray absorption
spectroscopy (SXAS) measurements were carried out at BL-1 beamline
of Indus-2 synchrotron radiation source at RRCAT, Indore. To study
the magnetic properties, magneto optical-Kerr effect (MOKE) and
polarized neutron reflectivity (PNR) measurements were carried
out. MOKE measurements were performed in longitudinal mode on M/S
Evico Magnetics system and PNR measurements were carried out at
the SuperADAM instrument at ILL, Grenoble, France.

\section{Results}
\label{res}
\subsection{\textbf{Composition and Structural Characterization}} \label{stru_char}

Nitrogen concentration of samples was measured using SIMS depth
profiles of Co and N (not shown). A \tcn~thin film sample with a
known nitrogen concentration of about 20 \pat~was used as a
reference. Using a procedure described in
ref.~\cite{CoN_AIP_Adv2015}, the nitrogen concentration was
calculated. We find that N concentration is about
5($\pm$2)\pat~for samples deposited using \pn = 25, 50, 75 and
100{\p} (table~\ref{table1}). This is surprising as it is
generally expected that with an increase in \pn, N concentration
should increase. This issue will be discussed in
section~\ref{diss}.

\begin{figure}\center
\includegraphics [width=90mm,height=70mm] {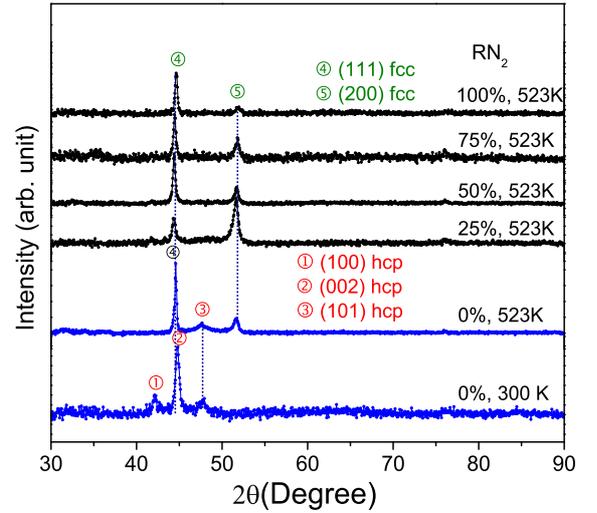}
\caption{\label{fig:xrd} (Color online) XRD pattern of Co-N thin
films deposited for various \pn = 0, 25, 50, 75 and 100{\p}~at
\Ts~of 523\,K.} \vspace{-5mm}
\end{figure}

As mentioned in section~\ref{expe}, we deposited two types of
sample, Co-N thin films with \pn = 25, 50, 75 and 100{\p} at \Ts =
523\,K~and pure Co films at \Ts = 300\,K~and 523\,K. Pure Co films
were deposited as a reference sample. The XRD pattern of all
samples is shown in fig.~\ref{fig:xrd}. Pure Co film deposited at
300\,K~show distinct peaks appearing at 2$\theta$ values
42.20$^{\circ}$, 44.75$^{\circ}$ and 47.51$^{\circ}$. Typical
error in measurements of peak position is about 0.06\,$^{\circ}$.
Comparing the position of these peaks with the standard JCPDS
reference number (050727), it can be inferred that these peaks
correspond to hcp Co planes (100), (002) and (101), respectively.
When the \Ts~is raised to 523\,K, peaks appearing in the XRD
pattern show a different behavior. We find that peaks now appear
at 2$\theta$ = 44.55$^{\circ}$, 47.64$^{\circ}$ and
51.57$^{\circ}$, which correspond to a mixture of fcc and hcp Co.
While peaks at 2$\theta$ = 44.55$^{\circ}$ and 51.57$^{\circ}$
correspond to fcc planes (111) and (200), respectively, the one
appearing in the middle is similar to the sample deposited at
300\,K.

\begin{figure}\center
\includegraphics [width=90mm,height=80mm] {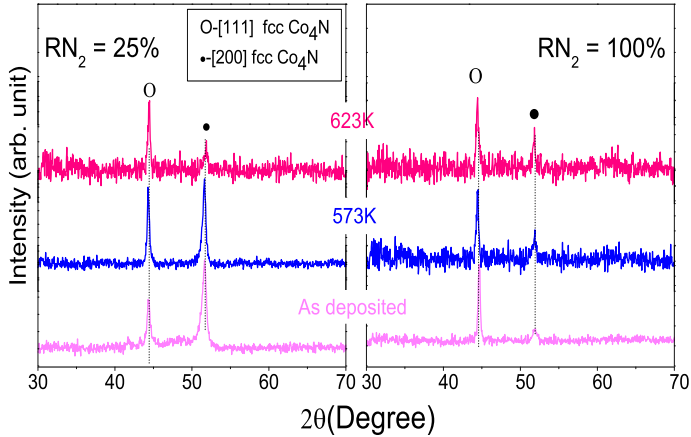}
\caption{\label{fig:xrd_annl} (Color online) XRD pattern of Co-N
thin film deposited at \Ts = 523\,K~for \pn = 25 and 100{\p} with
different annealing temperatures.} \vspace{-5mm}
\end{figure}

Addition of nitrogen gas during deposition seems to cause visible
changes in the XRD pattern (fig.~\ref{fig:xrd}). We find that the
peak corresponding to (002) plane of hcp Co does not appear
anymore. Additionally, the intensity of (111) is more than (200)
plane for samples having \pn = 50, 75, 100{\p} while for \pn =
25{\p} the intensity of (200) plane is more than (111) plane. In
addition, peaks corresponding to fcc planes (111) and (200) shift
towards lower 2$\theta$~values indicating an expansion in the
lattice parameter (LP). Such expansion may take place due to
interstitial occupation of N atoms (in fcc Co) and formation of
\tcn~phase. Calculated values of LP are shown in
table~\ref{table1}.

Moreover, we investigated the thermal stability of all samples
deposited at various \pn. Samples were annealed altogether in a
vacuum furnace with base Pressure of 1$\times$10$^{-6}$ Torr and
the annealing were carried out at 573\,K~and 623\,K~for 1 hour.
Representative XRD pattern of annealed samples for \pn = 25 and
100{\p} are shown in fig.~\ref{fig:xrd_annl}. We find that the
peak positions corresponding to (111) and (200) planes remain
identical. However the intensity of (111) plane seems to fluctuate
rather randomly, whereas no significant changes can be observed
for \pn = 100{\p} with annealing. From this it is clear that the
sample deposited at \Ts = 523\,K~remain stable when annealed upto
623\,K.

\begin{table} [!hb] \vspace{-5mm}
\caption{\label{table1} Parameters, \pat~of nitrogen, lattice
parameter, number density and average magnetic moment per Co
atom for Co-N thin film samples prepared at \Ts = 523\,K~for various
{\pn}.}
\begin{tabular}{ccccc} \hline
\pn&nitrogen&LP&$\rho$&magnetic\\
(\p)&\pat &\AA&(1e$^{28}$/m$^3$)&moment($\mu_\mathrm{B}$)\\
&($\pm$2)&($\pm$0.004)&($\pm$0.1)&($\pm0.05$)\\\hline\hline
0&0&3.521&8.9&1.68\\
25&4&3.536&8.7&1.75\\
50&5&3.536&8.7&1.70\\
75&5&3.530&8.7&1.73\\
100&6&3.532&8.7&1.7\\ \hline
\end{tabular}
\end{table}

\begin{figure}\center
\includegraphics [width=115mm,height=90mm] {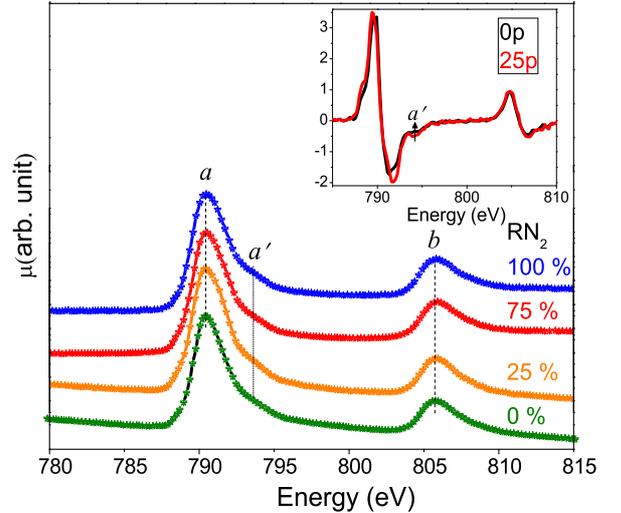}
\caption{\label{fig:xas} (Color online) XAS pattern of Co-N thin
films for \pn = 0, 25, 75 and 100{\p} at Ts = 523\,K~at Co L edge.} \vspace{-5mm}
\end{figure}

To know about the local environment and the electronic structure
of deposited films, XAS measurements were performed for samples
\pn = 0, 25, 75 and 100{\p} at Co L edges as shown in
fig.~\ref{fig:xas}. Inset to this figure compares the derivative
of absorption spectra of samples for \pn = 0 and 25{\p}. We find
three features assigned as $a$, $a^\prime$ and $b$. Features $a$
and $b$ have the energy difference of about 15\,eV~and are known
as L$_3$ (2p$_{3/2}$) and L$_2$ (2p$_{1/2}$) edge jumps, arising
due to well-known spin-orbit interaction. It can be noticed that
the position of features $a$ and $b$ remains identical across all
samples indicating that the oxidation state does not change with a
variation of \pn. In addition, the feature $a^\prime$, a shoulder
to feature $a$, shows a clear variation for samples deposited with
or without nitrogen. This feature appears due to different
chemical bonding at inequivalent 3d metal sites
~\cite{APL_11_K_Ito_Co4N, JAP_15_K_Ito_XAS} and is a
characteristic feature for anti-pervoskite type structures. We
also notice that the intensity of feature $a$, $a^\prime$ and $b$
is more (than pure Co) when nitrogen gas is used during
sputtering. This happens due to the enhanced hybridization N 2p
states with Co 3d states. Our XAS results clearly show a small but
observable difference between sample prepared with and without
nitrogen gas used during sputtering. However among the sample
sputtered with nitrogen (\pn = 25, 75 and 100\p), there seems to
be no observable difference in the XAS spectra.

\subsection{\textbf{Magnetic Measurements}} \label{mag_mes}

\begin{figure}\center
\includegraphics [width=100mm,height=80mm] {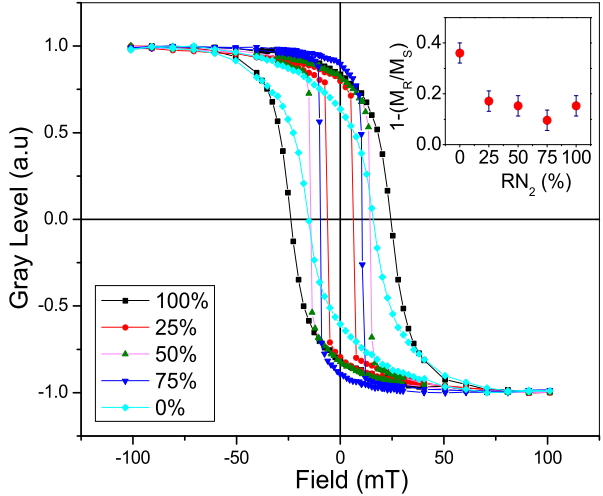}
\caption{\label{fig:moke} (Color online) Moke hysteresis loop of
Co-N thin films deposited at Ts = 523\,K for \pn = 0, 25, 50, 75 and
100{\p}; Inset shows the variation in 1-M$_R$/M$_S$ with respect
to {\pn}.}
\vspace{-5mm}
\end{figure}

Fig.~\ref{fig:moke} shows the MOKE hysteresis loops, typical of a
ferromagnetic sample. From the loops it can be seen that the
magnetic anisotropy (1-(M$_\mathrm{R}$/M$_\mathrm{S}$);
M$_\mathrm{R}$: remanent magnetization and M$_\mathrm{S}$:
saturation magnetization) and coercivity for pure Co and Co-N
films are different. Inset of the fig.~\ref{fig:moke} compares the
variation of 1-(M$_\mathrm{R}$/M$_\mathrm{S}$) with \pn. Here it
can be seen that the pure Co film has somewhat large anisotropy
than nitrogenated samples. While going from \pn = 25 to 100{\p}
only a little variation can be seen in
1-(M$_\mathrm{R}$/M$_\mathrm{S}$). It may noted that hcp Co is
expected to show rather large anisotropy but in fcc structure the
anisotropy becomes low (as low as 10 times than hcp). From our XRD
measurements, we found that the pure Co film deposited at \Ts =
523\,K~forms in a mixture of hcp and fcc phases, later being a
dominant phase. Therefore the magnetic anisotropy of Co is
expected to be low and it becomes even smaller when only fcc phase
get formed in presence of nitrogen gas during sputtering. Our
results obtained from XRD and XAS measurements correlate very well
to the fact that the pure Co sample is different from Co-N samples
and within Co-N samples they are identical.

\begin{figure}\center
\includegraphics [width=115mm,height=80mm] {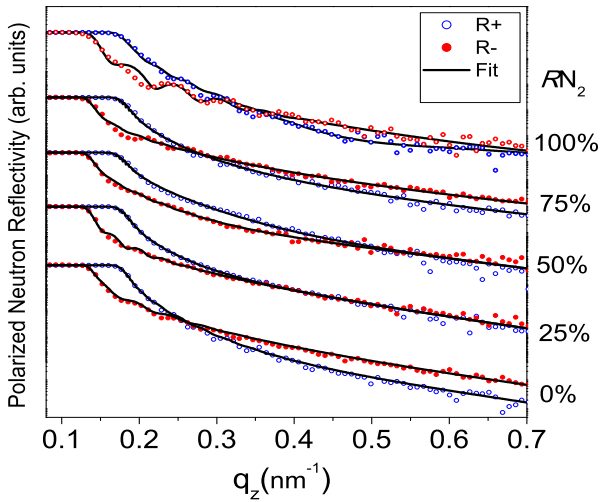}
\caption{\label{fig:pnr} (Color online) PNR pattern of Co-N thin
films for \pn = 0, 25, 50, 75 and 100{\p}.} \vspace{-5mm}
\end{figure}

In order to measure the absolute Ms of our samples, we did PNR
measurement. It is well-known that the PNR is the most accurate
method for magnetization measurements in thin films. Since PNR
provide an absolute value of Ms independent on sample mass and
does not get influence by diamagnetism of the substrate. We
performed the PNR measurements with an magnetic field (parallel to
sample surface) of 0.5\,Tesla to saturate them magnetically. The
PNR pattern (fig.~\ref{fig:pnr}) of the samples shows a clear
splitting between spin-up and down reflectivities which is a
signature of ferromagnetism. Here, it is interesting to see that
spin-up and spin-down reflectivities cross each other at different
q$_z$ values. This happen due to a thin surface layer
($\sim$3\,nm) of different density. Such layer is generally form
in thin films when exposed to atmosphere. To get the precise
information about the Ms and the number density, PNR data were
fitted using SimulReflec programme~\cite{SimulReflec}. Fitted
parameter are given in table~\ref{table1}. We find that pure Co
has number density 8.9$\times$10$^{28}$\,m$^{-3}$ and a Ms of 1.68
$\pm$ 0.05 $\mu_\mathrm{B}$ which is in agreement with bulk
Co~\cite{1986_E.P.Wohlfarth, TSF14_Silva}. When nitrogen is added,
we find number density decreases but magnetic moment increases
slightly. The obtained values of magnetic moments and number
density are shown in table~\ref{table1}.

\section{Discussion} \label{diss}
From the results presented in section \ref{res}, following
information can be obtained. (i) pure Co grows in a hcp structure
when deposited at \Ts~= 300\,K~whereas at \Ts~= 523\,K~a mixture of hcp
and fcc phases is formed. (ii) when nitrogen is added during deposition, a fcc
structure is formed, irrespective of the amount of nitrogen used
during sputtering. (iii) properties of pure Co and that of Co-N
films (for any \pn) are different with respect to structural,
electronic and magnetic properties.

As far as pure Co film deposited at \Ts~= 523\,K~is concerned, appearance of fcc Co phase already
at this \Ts~is somewhat surprising as the transition
temperature from hcp to fcc structure for bulk Co is 690\,K~or
above~\cite{cabral_1993_Co, 1999_Co_H_Zhang, CoN_AIP_Adv2015}.
However in case of thin films, it has been observed that such
transition from hcp to fcc phase can occur at much lower (than
bulk) transition temperature (690\,K) as observed by Carbel $et. al.$~\cite{cabral_1993_Co, 1999_Co_H_Zhang}. On the other hand, the
behavior of the films deposited in the presence of nitrogen is
surprising as an increase in \pn~is expected to increase in N
\pat. Generally for other systems like Fe-N, it has been observed
that the addition of nitrogen during deposition results in
N\pat~enhancement almost up to mononitride composition (1:1 for
Fe:N). Similarly, when Co-N films are deposited at \Ts~=
300\,K~they show an evolution of phases, starting with few \pat~of
N to 50\pat~N as \pn~increases from 5 to 100
{\p}~\cite{CoN_AIP_Adv2015}.

Therefore, our observation that independent of \pn~(at \Ts = 523\,K)
used, the phases form are always similar, is surprising and such
behavior has not been seen before for the Co-N system. On the
other hand a lot of experimental reports are available, claiming
the formation of \tcn~phase at \Ts~varying from 433\,K to
723\,K~\cite{JMS87_Oda, 2011_Co4N_K_Ito, JAC14_Lourenco,
Silva2015, TSF14_Silva}. Use of such higher \Ts~might have been
influenced by \tfn~phase which is generally formed at
\Ts$\sim$673\,K.

A comparison between the theoretical and experimentally observed
values of LP can be used to understand the results obtained in
this work. From our XRD data, we find a slight increment in the LP
of Co-N samples as compared to the pure Co film deposited at 523\,K.
Theoretical values of LP for fcc Co and \tcn~are 3.54\,{\AA} and
3.74\,{\AA}, respectively ~\cite{Silva2015, PRB07_Matar,
JMMM10_Imai}. This implies that for \tcn~(at N 20\pat)
composition, the lattice dilation should be about 5\p. However,
the experimental data available for \tcn~films shows that the
value of LP are 3.586\,{\AA} at 433\,K~\cite{JMS87_Oda},
3.524\,{\AA} at 523\,K~\cite{Silva2015}, 3.524\,{\AA} at
723\,K~\cite{2011_Co4N_K_Ito} etc. This implies that the dilation in
the LP of \tcn~phase obtain experimentally so far is $\sim$0.5 to
1\p~only, against the expected value of $\sim$5\p. However, when
\tcn~films are deposited at \Ts~= 300\,K, the measured value of LP
is 3.7\,{\AA}, quite close to the theoretical LP of
\tcn~\cite{CoN_AIP_Adv2015}.

Our data presented in this work, also imply that an expansion of
typically 0.5\p~in the LP would correspond to 2-3 N \pat~(in agreement with SIMS measurements).
This result clearly indicate that, when Co-N films are deposited at \Ts$\geq$300\,K, N
\pat~in resulting Co-N films is significantly low as compared to
those obtained in Fe-N system. The reason for such behavior can be
investigated while looking in the thermodynamics of Co-N system.
It is known that the enthalpy of formation ($\Delta$H$^{\circ}_f$) for
\tfn~is -12.2($\pm$20)\,kJ/mol~\cite{Tessier_SSS00}, however the
value of $\Delta$H$^{\circ}_f$ for Co-N system are not yet known
explicitly. In theoretical calculation carried by H\"{a}glund $et.~al.$,~\cite{1993_PRB_Haglund}, it has been shown that the
$\Delta$H$^{\circ}_f$ increases as one moves along 3d series from
Ti to Ni which means that the thermal stability of metal nitrides
should decreases as Z increases in the 3d
series~\cite{1993_PRB_Haglund}. In our recent study, we measured the
thermal stability of \tcn~films formed at \Ts~= 300\,K. Here it was
found that above a temperature of 423\,K, significant nitrogen
diffusion starts to take place. It seems at high \Ts, rapid N
diffusion takes place leaving behind pure Co with together with a
reminiscent fraction of \tcn. It is interesting to see that
fcc phase is formed for any \pn~used during sputtering, which also
indicate that N atoms diffuse out form a parent fcc \tcn~phase. In
addition since samples deposited at \Ts~= 523\,K~have somewhat
higher Ms than pure Co, and they are also stable upto 623\,K, is
an interesting observation that can be used to prepare fcc Co
films. Obtained results provide an important information about the
Co-N system and a recipe of preparation of \tcn~films which should
involve low \Ts~instead of higher temperature used in other works.

\section{Conclusion}
\label{con}

In this work we did a systematic study of Co-N thin films
deposited at \Ts~= 523\,K using \pn~= 0, 25, 50, 75 and 100{\p}. From our
SIMS, XRD, XAS, MOKE and PNR measurements, we found a clear
difference between samples deposited with or without nitrogen
during sputtering. However a variation in \pn~seems to cause not
much difference among samples even as \pn~is varied from 25 to
100{\p}. This is a counter intuitive result which can be understood
as the $\Delta$H$^{\circ}_f$ for the \tcn~might be much larger,
resulting in poor thermal stability leading to rapid diffusion of
N atoms. In such a scenario when Co-N films are deposited at high
\Ts, N atoms diffuse out, leaving behind a fcc Co phase which has
been often mistaken for a fcc \tcn~phase.

\section*{Acknowledgments}
We are thankful to V. R. Reddy, Zaineb Hussain for MOKE
measurements, Alexei Vorobiev for providing help in PNR
measurements, Layanta Behera and Anil Gome are acknowledge for
their help in various measurements. BL01, Indus 2 team (D. M.
Phase, D. K. Shukla, Rakesh Sah) are acknowledge for support in XAS
beamline.

\section*{References}

%\bibliography{TMN}

\end{document}